\documentclass[twocolumn,superscriptaddress,aps,prb,floatfix]{revtex4-1}

\usepackage{graphicx}
\usepackage{dcolumn}
\usepackage{bm}
\usepackage{color}
\usepackage[caption=false]{subfig} 

\usepackage{amsthm}
\usepackage{amsmath}
\usepackage{amssymb}

\newcommand{\Htc}{H_{\mathrm{TC}}}
\newcommand{\figref}[1]{Fig. \ref{#1}}

\begin{document}


\title{Relaxation dynamics of the toric code in contact with a thermal reservoir: finite-size scaling in a low temperature regime}

\author{C. Daniel Freeman}
\email{daniel.freeman@berkeley.edu}
\affiliation{Berkeley Quantum Information \& Computation Center, University of California, Berkeley, CA 94720, USA}
\affiliation{Department of Chemistry, University of California, Berkeley, CA 94720, USA}

\author{C. M. Herdman}
\altaffiliation[Present Address: ]{Institute for Quantum Computing, University of Waterloo, Waterloo, ON N2L 3G1, Canada}
\affiliation{Department of Physics, University of Vermont, Burlington, VT 05405, USA}

\author{Dylan Gorman}
\affiliation{Department of Physics, University of California, Berkeley, CA 94720, USA}

\author{K. B. Whaley}
\affiliation{Berkeley Quantum Information \& Computation Center, University of California, Berkeley, CA 94720, USA}
\affiliation{Department of Chemistry, University of California, Berkeley, CA 94720, USA}

\date{\today}

\begin{abstract}
 We present an analysis of the relaxation dynamics of finite-size topological qubits in contact with a thermal bath. Using a continuous-time Monte Carlo method, we explicitly compute the low-temperature nonequilibrium dynamics of the toric code on finite lattices. In contrast to the size-independent bound predicted for the toric code in the thermodynamic limit, we identify a low-temperature regime on finite lattices below a size-dependent crossover temperature with nontrivial finite-size and temperature scaling of the relaxation time. We demonstrate how this nontrivial finite-size scaling is governed by the scaling of topologically nontrivial two-dimensional classical random walks. The transition out of this low-temperature regime defines a dynamical finite-size crossover temperature that scales inversely with the log of the system size, in agreement with a crossover temperature defined from equilibrium properties. We find that both the finite-size and finite-temperature scaling are stronger in the low-temperature regime than above the crossover temperature. Since this finite-temperature scaling competes with the scaling of the robustness to unitary perturbations, this analysis may elucidate the scaling of memory lifetimes of possible physical realizations of topological qubits.
\end{abstract}

\maketitle

\section{Introduction}

The fact that the potential power of a large scale quantum computer has not yet been realized experimentally is due largely to the fragility of quantum information.  A ``conventional" quantum computer stores quantum information in spatially localized qubits--consequently local noise can generate errors that destroy the locally stored quantum information. The theoretical possibility of a fault tolerant quantum computer is well understood in the literature; in general this requires building redundancy into the experimental systems such that errors can be detected and corrected.  Although fault tolerance via such active error correction is theoretically feasible, the overhead required to perform active error correction has thus far kept a large scale quantum computer out of reach.

An alternative approach to fault-tolerant quantum computing is based on building \emph{physically} robust quantum hardware with passive error correction. The notion of a topological quantum computer builds on the possibility of storing quantum information nonlocally in a robust quantum phase of matter with topological order~\cite{Kitaev2003,Freedman2001,Wen1990,Nayak2008}; consequently, these phases of matter appear to hint at the potential design of a \emph{self-correcting} quantum computer. Indeed, while in equilibrium with a zerotemperature reservoir, a topological qubit is ``topologically protected,'' in that errors due to local perturbations are suppressed exponentially in the system size. Despite this promise, subsequent work has demonstrated that topological phases in two dimensions (2D) are thermally fragile because topological order is destroyed at any nonzero temperature~\cite{Castelnovo2007a,Nussinov2008a,Hastings2011a,Mazac2012,Iblisdir2010b}. While higher-dimensional topological phases are robust at finite temperature~\cite{Chamon2005,Castelnovo2008,Alicki2008,Bravyi2011b,Bravyi2013,Haah2011,Bacon2006a,Hastings2014}, it is only in 2D that such phases can act as a universal quantum computer based on topologically protected operations~\cite{Kitaev2003,Freedman2002a,Mochon2004b,Mochon2003,Nayak2008}. This shortcoming seems to preclude the possibility of a universal topologically protected quantum computer. 

Accordingly, the 2D toric code fails to be a true fault-tolerant quantum memory in 2D~\cite{Kitaev2003,Dennis2002,Bravyi1998,Alicki2007a,Alicki2009a,Bravyi2009,Yoshida2011b,Chesi2010b,Landon-Cardinal2013,Castelnovo2011a}. However, while topological order is destroyed at any finite temperature in the thermodynamic limit, on a finite-size system the topological order of the toric code nevertheless persists up to a finite-size crossover temperature~\cite{Castelnovo2007a}. This suggests the possibility of operating a topological qubit in a low-temperature regime where topological order persists due to finite-size effects. While finite-size effects reduce the zero-temperature robustness to unitary perturbations\cite{Kitaev2003}, the existence of a lowtemperature regime below the crossover temperature suggests that such finite-size effects may \emph{increase} the thermal robustness. Consequently, characterizing how the memory lifetime of a topological qubit depends on finite-size effects, especially in the low-temperature regime, is of practical importance.

In this paper, we use real-time Monte Carlo simulations to study the relaxation dynamics of finite-size topological qubits defined by the toric code, in contact with a thermal reservoir. Previous work using related methods focused on the high-temperature scaling of decoherence times~\cite{Chesi2010a,Rothlisberger2012,Al-Shimary2013,Hutter2012}; here we focus on the dynamics at low temperatures. We find a low-temperature regime that is well described by thermal relaxation dominated by quasiparticle pairs undergoing topologically nontrivial random walks. At higher temperatures, the decoherence is dominated by local creation and annihilation of quasiparticle pairs. The transition between these two regimes allows for a \emph{dynamical} definition of the crossover temperature $T^*$. We find that $T^* \sim 1/ \ln N$, which agrees with the scaling of a transition temperature defined from the topological entanglement entropy at equilibrium\cite{Castelnovo2007a}. Additionally we find that both the finite-size and finite-temperature scaling are stronger below than above $T^*$.

The structure of this paper is the following: in Sec. \ref{sec:TC} we present the relevant background of the toric code; Sec. \ref{sec:TCdynamics} introduces a microscopic master equation of the toric code interacting with a bath as well as an effective model of the low-temperature dynamics; Sec. \ref{sec:MC2DRW} presents a numerical study of topologically nontrivial random walks on a torus that we use to construct the low-temperature effective model; and, finally, Sec. \ref{sec:MCmethod} presents a numerical study of the microscopic master equation for the toric code interacting with a bath and an analysis of these results in comparison with the low-temperature effective model.

\section{The toric code}
\label{sec:TC}

\subsection{The toric code Hamiltonian}

\begin{figure}
\begin{center}
\scalebox{1}{\includegraphics[width=0.5\columnwidth]{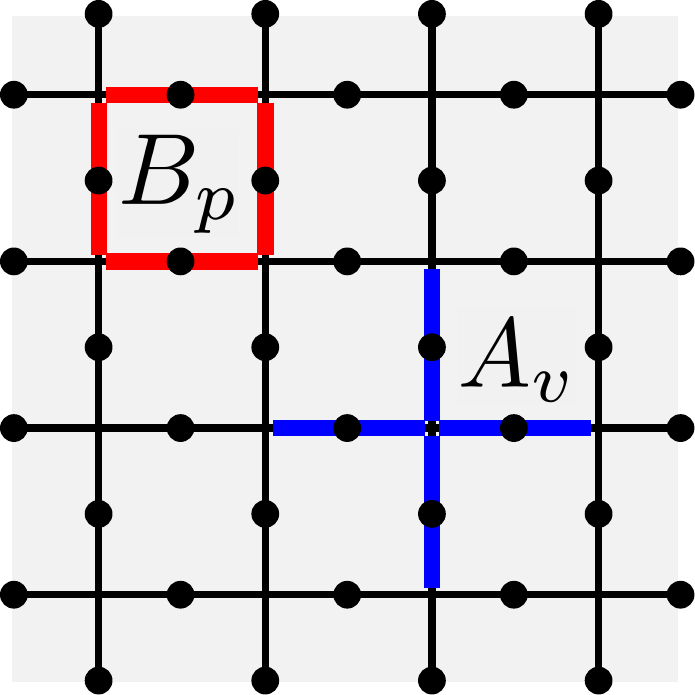}}
\end{center}
\caption{The vertex ($A_v$) and plaquette ($B_p$) operators of the toric code as defined in \eqref{eq:AvBp}.  Edges marked in red are operated on by $\sigma^x$ while those marked in blue are operated on by $\sigma^z$.}
\label{fig:TCoperators}
\end{figure}

The toric code provides a simple exactly soluble model with a topologically ordered ground state that may provide topologically protected qubits at $T=0$~\cite{Kitaev2003,Dennis2002}. The toric code is defined on a square lattice, where Ising spins sit on the links of the lattice. We define the linear dimension of the lattice as $L$ and the number of spins $N=2L^2$. The Hamiltonian involves four-spin interactions around the plaquettes and vertices of the lattice:
\begin{align}
\Htc &= -J_e \sum_v A_v -J_m \sum_p B_p ,\label{eq:HTC}\\
A_v &\equiv \prod_{j \in v} \sigma_j^z,\quad B_p \equiv \prod_{j \in p} \sigma_j^x,\label{eq:AvBp}
\end{align}
where the sums over $v$ and $p$ are over the vertices and plaquettes of the lattice, respectively (see \figref{fig:TCoperators}). The ground states are the $+1$ eigenstate of all $A_v$ and $B_p$ operators, since all such operators commute. On a torus, there are four degenerate ground states that are distinguished by the expectation values of non-local winding operators $W_{1,2}^{x}$,$W_{1,2}^{z}$:
\begin{align}
W_{1,2}^{x} \equiv \prod_{j \in \Gamma_{1,2}} \sigma_j^x,\quad W_{1,2}^{z} \equiv \prod_{j \in \tilde{\Gamma}_{1,2}}  \sigma_j^z,
\end{align}
where $\Gamma_{1,2}$  and $\tilde{\Gamma}_{1,2}$ are topologically non-trivial loops along the links and plaquettes of the lattice, respectively, that wind around each of the two axes of the torus. There is a finite gap $\Delta_{e,m} = 4J_{e,m}$ to excited states that are $-1$ eigenstates of some $A_v$ and/or $B_p$.  These correspond to $e$-type and $m$-type \emph{anyonic} quasiparticle excitations, respectively \cite{Kitaev2003}.

\subsection{The toric code as a quantum memory}

Consider the $W_{1,2}^z$ basis for the degenerate ground state subspace; we may label the four ground states by the eigenvalues of  $W_{1,2}^z$:
\begin{equation}
\Bigl \{ \left \vert \Psi_0 \right \rangle \Bigr \} = \Bigl \{ \left \vert \Psi_0^{++} \right \rangle, \left \vert \Psi_0^{-+} \right \rangle, \left \vert \Psi_0^{-+} \right \rangle, \left \vert \Psi_0^{--} \right \rangle\Bigr \}, 
\end{equation}
where $+/-$ represent the $\pm1$ eigenstates of $W_{1}^z$ and $W_{2}^z$, respectively. We choose this basis to be the logical basis for a two qubit quantum memory. To simplify the discussion of errors we will consider the limit $J_m \rightarrow \infty$, so that only $e$-type quasiparticles have finite energy. The lowest excited eigenstates have a single pair of localized $e$-type quasiparticles which are connected to a ground state by operation of an error string:
\begin{equation}
\left \vert e_v, e_{v'} \right \rangle = S_x \bigl[  \Gamma_{v,v'} \bigr]  \bigl \vert \Psi_0 \bigr \rangle, \quad  S_x \bigl[  \Gamma_{v,v'} \bigr] \equiv \prod_{j \in \Gamma_{v,v'}} \sigma_j^x.
\end{equation}
Here, $v$ and $v'$ are the vertices where the quasiparticles are located and the string $\Gamma_{v,v'}$ connects $v$ and $v'$. Error strings that form topologically non-trivial loops generate the $W_{1,2}^{x}$ operators and drive transitions between ground states; such error strings create {\it noncorrectable errors}, i.e., errors in the logical subspace that cannot be corrected.  {\it Correctable errors}, or self-correcting errors, are those error strings that close without causing a change in winding number.  

Under local perturbations to $\Htc$, such topologically nontrivial error strings only occur at order $L$ in perturbation theory; consequently both the splitting of the ground state degeneracy and transitions between ground states are suppressed exponentially, and thus this ground state subspace is ``topologically protected" from such unitary perturbations~\cite{Klich2010,Bravyi2010a,Bravyi2011a,Trebst2007,Tupitsyn2010,Dusuel2011,Michalakis2013,Kay2011a}. The toric code can thus act as a self-correcting quantum memory at \emph{zero} temperature.

\subsection{The toric code at finite temperatures}

Despite the topological protection at zero temperature, in the thermodynamic limit the topological order of the toric code is destroyed at any finite temperature~\cite{Nussinov2008a,Castelnovo2007a}. Consequently, a topological qubit would be thermally fragile. While topologically nontrivial error strings due to {\it unitary} perturbations are exponentially suppressed in the system size, non-trivial error strings may also be generated by non-unitary perturbations, e.g., from contact with a thermal reservoir. Non-correctable errors due to non-unitary perturbations are not exponentially suppressed in the system size.

The thermal fragility of the 2D toric code can be understood from a simple picture of the dissipative dynamics generated from local interactions with an external bath. A local system-bath interaction can generate a {\it trivial error string} by flipping a single spin, thus creating a single pair of neighboring quasiparticles:
\begin{equation}
\sigma_j^x \left \vert \Psi_0 \right \rangle = \left \vert e_v, e_{v'} \right \rangle
\end{equation}
where $v$ and $v'$ are the vertices on either end of the link $j$. The rate of such a process is suppressed exponentially in the inverse temperature, due to the energy gap $\Delta$ to such excited states. Such a trivial error is correctable by applying another $\sigma_j^x$. However, additional trivial error strings $\sigma_{j'}^x$ with $j\neq j'$ applied to $v$ or $v'$ will generate a longer, {\it non-trivial error string} at no energy cost. Consequently, local coupling to a bath can drive a random walk of quasiparticle pairs around the lattice with a rate that is only suppressed by a single Boltzmann factor. Such random walks may generate topologically nontrivial error loops and return the system back to the ground state subspace. If an error loop has an odd winding number, this error loop has caused a non-correctable error by driving a transition between ground states. Alternatively, if the error loop has an even winding number, the error is self-correcting.  Indeed, Alicki {\it et al.} have placed a system-size-independent upper bound on the relaxation time of a pure toric code ground state that explicitly demonstrates this thermal fragility of the toric code in the thermodynamic limit~\cite{Alicki2007a,Alicki2009a}. Additionally, the analysis of Nussinov and Ortiz demonstrates the lack of ``spontaneous topological symmetry breaking" at finite-temperature, as the autocorrelation time of the winding operators is sub-exponential in lattice dimension in the thermodynamic limit~\cite{Nussinov2008a,Nussinov2009a}.

\subsection{Crossover temperature }

While the toric code is thermally fragile at all non-zero temperatures in the thermodynamic limit, we can also consider how this fragility is affected by the finite size of a lattice.  As outlined above, the dissipative error processes which lead to thermal fragility occur when there is a single quasiparticle pair present. Since the number of excitations in equilibrium is suppressed by the Boltzmann factor at low temperatures, we expect that at sufficiently low temperatures the number of quasiparticles in equilibrium will be vanishingly small. We can then define an {\it equilibrium crossover temperature} $T_{eq}^*$ which distinguishes the thermally fragile regime from a low temperature regime with reduced dissipative error processes by:
\begin{equation}
N e^{-\Delta/T_{eq}^*} \sim 1 \Rightarrow T_{eq}^* \sim \frac{\Delta}{\ln N}.
\end{equation} 

Castelnovo and Chamon define an equilibrium crossover temperature $T_{eq}^*$ above which the topological entanglement entropy vanishes~\cite{Castelnovo2007a}. They find that this equilibrium definition of $T^*$ scales inversely with the log of the system size, and that this  becomes a zero temperature phase transition in the thermodynamic limit. Conversely, on a finite sized system the crossover temperature $T^*$ defines a low temperature regime where topological order persists as a finite-size effect.  This opens the possibility of using finite-size effects to exploit the zero temperature topological order at finite temperatures. The usefulness of this low temperature regime for quantum information processing depends on the scaling of the relaxation time in this regime. Robustness to unitary perturbations requires a sufficiently large system size to minimize the splitting of the degeneracy and the matrix elements between ground states, while the thermal fragility increases with system size.  Thus one may expect that there is an optimal size for computational performance.

Below, we directly address this question of the finite-size scaling of the relaxation time of a toric code ground state in contact with a thermal reservoir.  This analysis complements the growing literature concerning \textit{active} error correction on the toric code by use of the stabilizer space and associated stabilizer operations~\cite{Dennis2002,Bravyi2009,Chesi2010b,Jouzdani2013}.  Usually, stabilizer error analysis is considered in the context of effectively infinite temperature thermal instability\cite{Chesi2010a}, in contrast to the finite-temperature dynamics presented here.  A complete toolkit for understanding and controlling errors in physical implementations of the toric code would need a predictive low temperature model, as well as a recipe for understanding how the fidelity of stabilizer operations affects the \textit{finite} temperature operation of the toric code.  Here we focus on the robustness of the passive error correcting (i.e., self-correcting) dynamics under the action of the toric code Hamiltonian in contact with a thermal reservoir.

\section{Dynamics of the Toric Code in contact with an external bath}
\label{sec:TCdynamics}

\subsection{Microscopic Quantum Master Equation}
\label{sec:TCME}

We present a microscopic model of the real-time non-equilibrium dynamics of the toric code in contact with a thermal reservoir. Due to the fact that the spectrum of $\Htc$ has a finite gap to excited eigenstates with localized quasiparticle excitations, such dynamics may be described by a Lindblad master equation\cite{Alicki2007a,Alicki2009a}:
\begin{align}
\dot{\rho }=\sum_{\omega }{2c_{\omega }\rho c^{\dagger }_\omega}-c^{\dagger }_{\omega }c_{\omega }\rho -\rho c^{\dagger }_{\omega }c_{\omega }, \label{eq:Lindblad}
\end{align}
here $\rho$ is the toric code system density matrix and $\{c_\omega\}$ is a set of Lindblad operators generated by local system-bath interactions. We will consider the limit where $J_m \gg J_e$, such that at low temperatures the system will remain in the +1 eigensector of all $B_p$ operators and only $e$-type quasiparticles will be excited by the reservoir. We will only consider local system-bath couplings, for which the bath generates single spin flips in the system. The relevant Lindblad operators are:
\begin{equation}
\left \{ c_\omega \right \} = \left \{ \sqrt{\gamma_0} T^e_{vv}, \sqrt{\gamma_+} E^{e\dagger}_{vv}, \sqrt{\gamma_-} E^{e}_{vv} \right\} 
\end{equation}
where $E^{e\dagger }_{vv'}$ ($E^{e}_{vv'}$) creates (annihilates) a pair of quasiparticles at neighboring vertices and $T^e_{vv'}$ translates a quasiparticles across a link.  These operators are defined by:
\begin{align}
E^{e\dagger }_{vv'}&=\frac{1}{4}{\sigma }^x_{vv'}\left(1-A_{v}\right)\left(1-A_{v'}\right),\notag \\ \label{eq:LindbladDef}
T^e_{vv'}&=\frac{1}{4}{\sigma }^x_{vv'}\left(1-A_{v}\right)\left(1+A_{v'}\right).
\end{align}

\begin{figure}
\begin{center}
\scalebox{1}{\includegraphics[width=0.5\columnwidth]{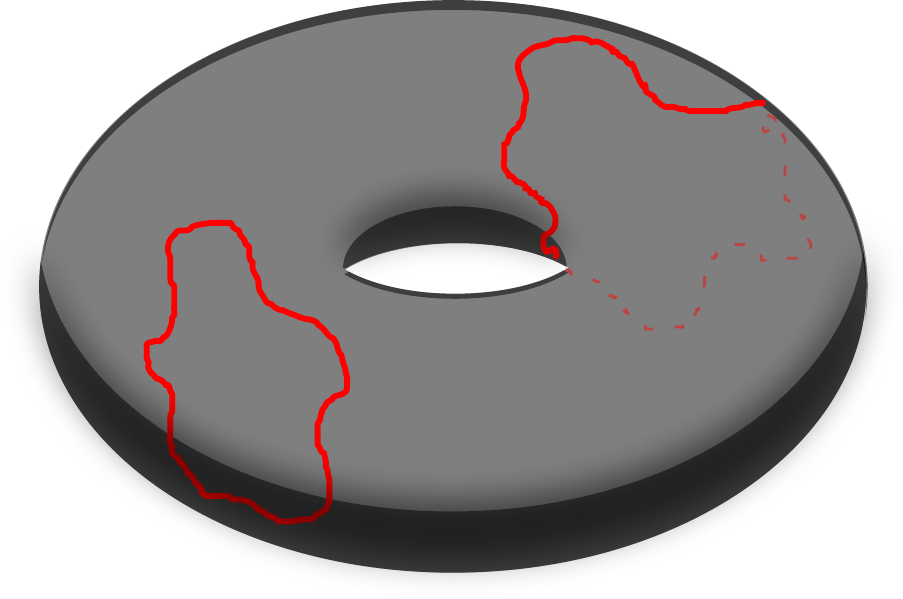}}
\end{center}
\caption{A torus with a self-correcting error string (left) and an uncorrectable error string (right).}
\label{fig:ToricCartoon}
\end{figure}

Since the Lindblad form of the master equation only connects diagonal elements of the density matrix $\rho$ to other diagonal elements, expectation values of diagonal elements will evolve independently of off-diagonal density matrix elements. Correspondingly, the time evolution of diagonal matrix elements reduces to a classical master equation:
\begin{align}
\frac{d P_n}{d t} = \gamma_0 \sum_{n_0}  \left( P_{n_0}-P_n \right)&+ \sum_{n_+}  \left( \gamma_- P_{n_+}- \gamma_+ P_n \right) \notag \\
&+ \sum_{n_-} \left( \gamma_+ P_{n_-} - \gamma_- P_n \right) \label{eq:ME}
\end{align}
where $n$ labels an eigenstate of $\Htc$, $P_n = \rho_{nn}$ are the diagonal matrix elements (probabilities) and $\{\gamma_0,\gamma_+,\gamma_-\}$ are the rates at which the operators $\{ T^e,E^{e\dagger},E^e\}$ act, respectively. Similarly, in \eqref{eq:ME}, for a given $n$, the indices of $n_0$, $n_+$, and $n_-$ label the sets of eigenstates connected to $|n\rangle$ by the operators $T^e$, $E^{e\dagger}$, and $E^e$, respectively. The ratio $\gamma_+/\gamma_-$ is fixed by detailed balance to be
\begin{equation}
\frac{\gamma_+}{\gamma_-} = e^{-\Delta/T} \label{eq:DetBal}
\end{equation}
but the nature of the bath and the coupling strength determines $\gamma_0$ and the magnitude of $\gamma_-$ (or equivalently $\gamma_+$). 

We consider here an Ohmic, Markovian bath with a power spectrum given by:
\begin{align}
J\left(\omega \right)=\omega e^{-\frac{\omega }{{\omega }_c}}
\end{align}
with ${\omega }_c$ a high frequency cutoff much larger than $J_e$.  Taking $w_c$ to infinity gives rise to decay rates of the form\cite{Chesi2010a}:
\begin{align}
\gamma \left( \omega \right)=\xi \left  \vert \frac{\omega }{1-e^{-\beta \omega }} \right \vert 
\end{align}
where $\xi $ sets the strength of the phenomenological system-bath coupling. This leads to the following rates:
\begin{equation}
\gamma_0 \equiv \frac{\xi }{\beta },\quad \gamma_+ = \frac{\xi \Delta }{e^{\beta \Delta }-1},\quad \gamma _-= \frac{\xi \Delta }{1-e^{-\beta \Delta }}. \label{eq:gamma}
\end{equation}

We are most interested in the dynamics deriving from the initial condition of a pure ground state. We characterize the relaxation from a pure ground state by considering the time evolution of the expectation value of the winding operators:
\begin{equation}
\left \langle W^Z_{1,2} \left( t \right) \right \rangle \equiv \mathrm{Tr} \left[ \rho \left( t \right) W^Z_{1,2} \right]. \label{eq:expW}
\end{equation}
The population dynamics are governed by the creation of quasiparticle pairs that undergo random walks on the torus and then annihilate. Thermal transitions between ground states occur when the quasiparticle pair undergoes a  {\it topologically non-trivial} random walk before annihilating.  The decay of the expectation values in \eqref{eq:expW} from their values in a pure state with eigenvalue $\pm1$ can be due to both topologically nontrivial random walks generating transitions between ground states, as well as transitions to excited states via propagating, open error strings. Consequently, the statistics of such topological random walks affect the scaling of the lifetime of a ground state.

\subsection{Comparison to Ising Model Dynamics}
\label{sec:Ising}

Nussinov and Ortiz showed that one can take advantage of the equivalence of the partition function of the toric code and that of 1D classical Ising chains to compute equilibrium properties of the toric code~\cite{Nussinov2008a,Nussinov2009a}.  However, one can not readily take advantage of this mapping for the study of non-equilibrium properties. While the partition function is only a function of the spectrum of the system, non-equilibrium dynamics depend on the nature of the (local) coupling to the external reservoir. Since the mapping of the toric code to an Ising chain maps a 2D model onto a 1D model, local couplings of the toric code to an external reservoir in general can lead to non-local couplings in the corresponding Ising model. Thus, a simple model of the 1D non-equilibrium dynamics of the Ising chains {\it locally} coupled to an external bath cannot describe the non-equilibrium dynamics of the toric code with a {\it local} bath coupling. Fundamentally, the dynamics of each system at low temperatures are governed by the random walk of defects (anyons in the toric code, domain walls in the Ising models); the defects of the Ising model undergo 1D random walks, whereas those of the toric code undergo 2D random walks. Consequently, we cannot directly compute the finite-size relaxation times of the toric code ground states from an analysis of the dynamics of the Ising chain. It is nevertheless useful to discuss the nature of thermal relaxation in a finite-sized Ising chain to help inform our discussion of such dynamics in the toric code. 

Consider a periodic 1D chain of $L$ classical Ising spins $s_i=\pm1$ with energy
\begin{equation}
E = -J \sum_i s_i s_{i+1}, \label{eq:Eising}
\end{equation}
where $J>0$ is a ferromagnetic coupling constant. The ground state of \eqref{eq:Eising} is a ferromagnet, but the long range order is destroyed at all nonzero temperatures. Excitations above the degenerate ground states are pairs of domain walls with energy cost $\Delta = 4J$. If a pair of domain walls undergoes a topologically non-trivial 1D random walk, this drives a thermal transition between ground states, which is we will refer to as ``ground state relaxation''. At sufficiently low temperatures on a finite-sized system, we can expect that these 1D topologically nontrivial walks will dominate the relaxation time of the magnetization. Such a low temperature regime must occur when there is less than a single pair of defects in equilibrium:
\begin{equation}
L \cdot e^{-\Delta/T} \ll 1.
\end{equation}
At low enough temperatures, there may be a separation of time scales such that $\gamma_+ \ll \gamma_0 \ll \gamma_ -$. Intuitively, the domain wall production rate, $\gamma_+$, can be tuned much less than the domain wall annihilation rate, $\gamma_-$, simply by lowering the temperature (c.f. \eqref{eq:DetBal}).  For certain choices of bath model, the domain wall hopping rate, $\gamma_0$, can be tuned between the latter two rates. On a finite size lattice the time scale for an extensive random walk of the defects can be estimated by the diffusion equation to be of the order of $L^2/\gamma_0$. We consider the low temperature regime of a finite size lattice where $\gamma_+ \ll \gamma_0/L^2$. In this limit, the rate of topologically nontrivial walks occurring is determined by the rate of production of defect pairs and by the probability that such pairs undergo a topologically nontrivial walk before annihilating.  This is because any domain wall pairs which proceed to an extensive random walk will carry out their walk and annihilate much faster than another pair of defects will be created.

We consider only the lowest order processes at first order in $\gamma_+$. Once a defect pair is created, the probability of the pair separating instead of trivially annihilating is of the order of $\gamma_0/\gamma_-$. The lowest order processes will annihilate upon their first return to neighboring links.  Processes for which the defect pair do not annihilate after the first return to neighboring links will occur at higher order in $\gamma_0/\gamma_-$. Consequently the overall order of these lowest order processes is $\gamma_+ \gamma_0 / \gamma_-$. We may then introduce a phenomenological form of the relaxation rate from a ferromagnetic ground state: 
\begin{align}
{\Gamma }_{\rm{Ising}}(\beta ,N) \sim \gamma_0 \cdot e^{-\Delta/T}  \cdot L \cdot P^{\Omega }_{{\rm 1D}} \left( L \right). \label{eq:IsingGamma}
\end{align}
The linear scaling in $L$ arises from the number of locations for domain wall pairs to be created. The factor $P^{\Omega }_{{\rm 1D}}(L)$ is the probability that any given domain wall that does not immediately annihilate eventually undergoes a topologically nontrivial random walk. A symmetry argument shows that $P^{\Omega }_{{\rm 1D}} (L )\sim L^{-1}$ (see Appendix \ref{sec:P1D}). This linear scaling of $P^{\Omega }_{{\rm 1D}} (L )$ suggests that the relaxation rate of the Ising model is size independent: ${\Gamma }_{\rm{Ising}} \sim \gamma_0 \cdot e^{-\Delta/T}$.

In Ref.~\onlinecite{Glauber1963d} Glauber solved the exact dynamics of the Ising chain in contact with a thermal reservoir. Glauber uses a bath where $\gamma_0$ is taken to be a constant and 
\begin{equation}
\frac{\gamma_0}{\gamma_-} = \frac{1}{2} \frac{1}{1-e^{-\Delta/T}}. 
\end{equation}
The relaxation time of this model is found to be
\begin{equation}
\Gamma_{\rm{Glauber}} = \frac{\gamma_0}{1+e^{\Delta/T}}
\end{equation}
At low temperatures, $\Gamma_{\rm{Glauber}} \approx \gamma_0 e^{-\Delta/T}$, in agreement with the expected size independent form of the low temperature single defect pair model\eqref{eq:IsingGamma}, despite the fact that $\gamma_0/\gamma_- \ll 1$ is not satisfied.

\subsection{Low temperature phenomenological dynamics}

We can now make an analogous argument for the toric code. At sufficiently low temperatures on a finite-sized lattice, the relaxation rate from a ground state should be dominated by the dynamics of a single quasiparticle pair. Transitions between ground states are generated by pairs of excitations that annihilate after undergoing a topologically non-trival 2D random walk with an odd winding number. The low temperature regime dominated by single defect pairs occurs when 
\begin{equation}
L^2 \cdot e^{-\Delta/T} \ll 1.
\end{equation}
We consider a separation of time scales for the Ohmic bath defined by \eqref{eq:gamma}:
\begin{align}
\gamma_0^{-1} L^2  \ll \gamma_+^{-1} &\Rightarrow L \ll \sqrt{\frac{T}{\Delta}} e^{\Delta/2T} \\
\gamma_-^{-1}  \ll \gamma_0^{-1} &\Rightarrow \frac{T}{\Delta} \ll 1
\end{align}
In this regime, the lowest-order processes are of the order of,
\begin{equation}
\gamma_0 \frac{\gamma_+}{\gamma_-} \sim \xi \cdot T \cdot e^{-\Delta/T}
\end{equation}
and we write a phenomenological ground state relaxation rate of the form,
\begin{align}
{\Gamma }_{\rm{TC}}(\beta ,L) \sim \xi \cdot T \cdot e^{-\Delta/T} \cdot L^2\cdot P^{\Omega }_{{\rm 2D}} \left( L \right). \label{eq:GammaTC}
\end{align}
Analogous to the phenomenological relaxation rate for the Ising model (i.e., \eqref{eq:IsingGamma}), the term $\xi \cdot T \cdot e^{-\Delta/T} \cdot L^2$ encodes the rate at which free quasiparticle pairs are produced.  The number of spins where a defect pair can be created is $N=2L^2$. The scaling of the topological factor $P^{\Omega }_{{\rm 2D}}(L )$, which is the probability of a 2D topologically nontrivial walk (i.e., a walk with odd winding number) will control finite-size scaling of $\Gamma_{\rm{TC}}$; only if $P^{\Omega }_{{\rm 2D}}(L ) \sim L^{-2}$ will the relaxation rate of the toric code be system size independent, as for the classical Ising chain.  Note that $P^{\Omega }_{{\rm 2D}} \left(L \right)$ is, in general, a function of temperature (see Appendix A).

\begin{figure}
\begin{center}
\scalebox{1}{\includegraphics[width=0.5\columnwidth]{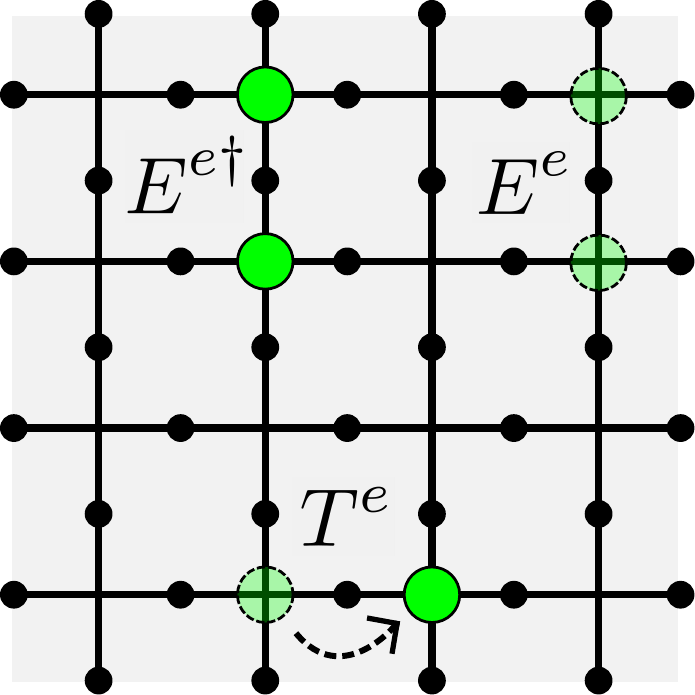}}
\end{center}
\caption{A depiction of the operation of the Lindblad operators in the master equation describing the interaction of the toric code with an external bath, as defined in \eqref{eq:LindbladDef}.}
\label{fig:Krausoperators}
\end{figure}

\subsection{High temperature phenomenological dynamics}

 At high temperatures, the relaxation of a toric code ground state will be dominated by the growing population of quasiparticles, rather than the dynamics of a single quasiparticle pair. In this regime, $\gamma_+ \approx \gamma_- \approx \gamma_0$ and the relaxation rate, i.e., the rate of decay of the expectation values $\langle W_{1,2}^z \rangle$, is due to error strings created across the length of the $ W_{1,2}^z$ operator. Consequently we expect the decay to be {\it linear} in system size:
 \begin{equation}
\Gamma_{T_H} \sim \gamma_+ L. \label{eq:GammaTH}
 \end{equation}
This linear scaling arises from the short time dynamics of the master equation \eqref{eq:Lindblad} \cite{Viyuela2012} and is independent of the topological processes that contribute to $P^{\Omega }_{{\rm 2D}}(L )$ and dominate the low temperature regime.

\subsection{Low temperature effective model of ground state transitions}

We will now use the form of the scaling of non-trivial annihilation probabilities discussed above to construct an effective low temperature minimal Markov model of the toric code ground state subspace. The ground state Markov model is defined by the master equation
\begin{equation}
\frac{d \boldsymbol{P}} {dt} = \boldsymbol{\Gamma} \boldsymbol{P} \left( t \right) \label{eq:Markov}
\end{equation}
where $\boldsymbol{P} = (P_{++},P_{+-},P_{-+},P_{--})$ is the vector of all ground state probabilities and $\boldsymbol{\Gamma}$ is the matrix of transition rates between ground states. 

Here we assume the low temperature form of the transition rate between ground states to take the form of \eqref{eq:GammaTC}; correspondingly the matrix elements of $\boldsymbol{\Gamma}$ take the form
\begin{equation}
\Gamma_{ij} = \lambda P_{ij}^{\Omega} 
\end{equation}
where $\Gamma_{ij}$ corresponds to the transition $i \rightarrow j$, $\lambda$ is a rate of production of anyon pairs that undergo a nontrivial random walk, and $P_{ij}$ is the probability that a given anyon pair will undergo a topologically nontrivial walk causing the transition $i\rightarrow j$.
We take the form of $\lambda$ to be:
\begin{align}
\lambda=2 L^2 \gamma_+ \left(1-\frac{\gamma_-}{6 \gamma_0 +\left(2 L^2-7\right)\gamma_+ + \gamma_-}\right),\label{eq:lowTModel}
\end{align}
where $2 L^2 \gamma_+$ is the rate of pair creation for the entire lattice.  The remaining factor in brackets is exactly the probability that an adjacent pair of quasiparticles on an otherwise empty lattice does not annihilate.  The numerical factors (i.e., $6, (2L^2-7), 1$) simply index the number of edges that can be acted upon by the different Lindblad operators for the lattice configuration with a single pair of adjacent quasiparticles.  By detailed balance, the probability of a given Lindblad operator acting on the system (e.g., $E^e$) is then just the ratio of the rate of that operator (e.g., $\gamma_-$) to the weighted sum of the rates of the other available operators, weighted by the number of edges available to each operator (e.g., $6 \gamma_0 +\left(2 L^2-7\right)\gamma_+ + \gamma_-$).  Thus, $\lambda$ accounts for the creation rate of quasiparticle pairs that do not immediately annihilate, or those pairs which can generate nontrivial random walks.

The form of $P_{ij}^{\Omega}$ is determined by whether the matrix element is relating ground states that differ by a winding on one axis of the torus or on both. We define $P^{\Omega}_{\delta 1}$ and $P^{\Omega}_{\delta 2}$ to be the probabilities of a topologically nontrivial annihilation that has an odd winding about one axis and both axes of the torus, respectively. Then the form of $P_{ij}$ is:
\begin{equation}
P_{ij}^{\Omega} = 
\begin{cases} P^{\Omega}_{\delta 1} &\mbox{if } i,j  \mbox{ differ by one winding number } \\ 
P^{\Omega}_{\delta 2} &\mbox{if } i,j  \mbox{ differ by both winding numbers } \\
- 2 P^{\Omega}_{\delta 1}-  P^{\Omega}_{\delta 2} &\mbox{if } i=j  
\end{cases} 
\end{equation}

We may solve this Markov model exactly by integrating \eqref{eq:Markov}: for $P_{++} (t=0) = 1$ we obtain:
\begin{align}
P_{++}\left(t\right) &= \frac{1}{4}(1+e^{-4tP^{\Omega}_{\delta 1} \lambda }+2e^{-2t\left(P^{\Omega}_{\delta 1} +P^{\Omega}_{\delta 2} \right)\lambda }) \notag \\
P_{+-}\left(t\right) &= P_{-+}\left(t\right)=\frac{1}{4}\left( 1- e^{-4tP^{\Omega}_{\delta 1} \lambda } \right) \notag \\
P_{--}\left(t\right) &=\frac{1}{4}(1+e^{-4tP^{\Omega}_{\delta 1} \lambda }-2e^{-2t\left(P^{\Omega}_{\delta 1} +P^{\Omega}_{\delta 2} \right)\lambda }).
\end{align}
When $P^{\Omega}_{\delta 1} \approx P^{\Omega}_{\delta 2} \approx P^{\Omega}_{\rm{2D}}$, we see that all $P(t)$ are well described by an exponential decay with rate $4 P^{\Omega}_{\rm{2D}} (L) \lambda$ at a finite temperature, $T$ (see Appendix \ref{sec:resum}).

\section{Topologically non-trivial random walks on the torus}
 \label{sec:MC2DRW}

\subsection{Scaling of topologically non-trival wallks}

As discussed above, at sufficiently low temperatures on a finite size lattice, we expect the relaxation time of a toric code ground state to depend on the statistics of topologically nontrivial random walks on the torus.  In this section we present a numerical study of discrete random walks on a square lattice on a torus using Monte Carlo simulations. Without loss of generality, we may map the processes of pair creation, two-particle random walk, and annihilation to a single random walker undergoing a random walk that starts and ends at the origin.  We can compute the probability of a quasiparticle pair generating a transition between ground states after annihilation from the statistics of topologically non-trivial walks of the single walker with odd winding.

To estimate the scaling of $P^{\Omega }_{{\rm 2D}}(L )$, we consider a related quantity: the probability that two random walkers will annihilate after $n$ steps $p(n)$. Topological walks must have radius of $L$; given that the radius of a 2D random walk scales as $\sqrt{n}$, we may assume that topological walks hve a minimum number of steps that scales as $n_{\rm{topo}} \sim L^2$. $P^{\Omega }_{{\rm 2D}}(L )$ may then be estimated as
\begin{align}
P^{\Omega }_{{\rm 2D}} \sim \int^{\infty}_{n_{\rm{topo}}} dn~p \left(n\right).
\end{align}
This rough estimate assumes that all walks larger than a certain length are necessarily topologically nontrivial.

Restricting to a planar square lattice with trivial topology, the annihilation probability $p^{\rm{p}}(n)$ can be computed
to give the asymptotic behavior for large $n$ as~\cite{Montalenti2000}:
\begin{equation}
p^{\rm{p}} \left(2n\right) \approx \frac{1}{2n{\left({\ln  2n\ }\right)}^2}. \label{eq:ppln}
\end{equation}
For small $n$, the exact result may be computed numerically via a recursion relation~\cite{Montalenti2000}. We then can estimate the scaling of $P^{\Omega }_{{\rm 2D}}$ by integrating the planar result:
\begin{align}
P^{\Omega }_{{\rm 2D}} \left( L \right) &\sim \int^{\infty }_{n_{\rm{topo}}} dn~ \frac{1}{ n_{\rm{topo}}  \left( \ln n_{\rm{topo}} \right)^2 }=\frac{1}{\ln n_{\rm{topo}} } \notag \\
 & \sim  \frac{1}{ \ln L }.
\end{align}
We expect then that the poly-log scaling of $p^{\rm{p}}(n)$ will lead to an inverse logarithmic finite-size scaling of $P^{\Omega }_{{\rm 2D}}$. Below we compute $P^{\Omega }_{{\rm 2D}}$ numerically and demonstrate this finite-size scaling empirically.

\subsection{Monte Carlo study of topologically non-trivial random walks on the torus}

\begin{figure}
\begin{center}
\scalebox{1}{\includegraphics*[width=0.5\columnwidth]{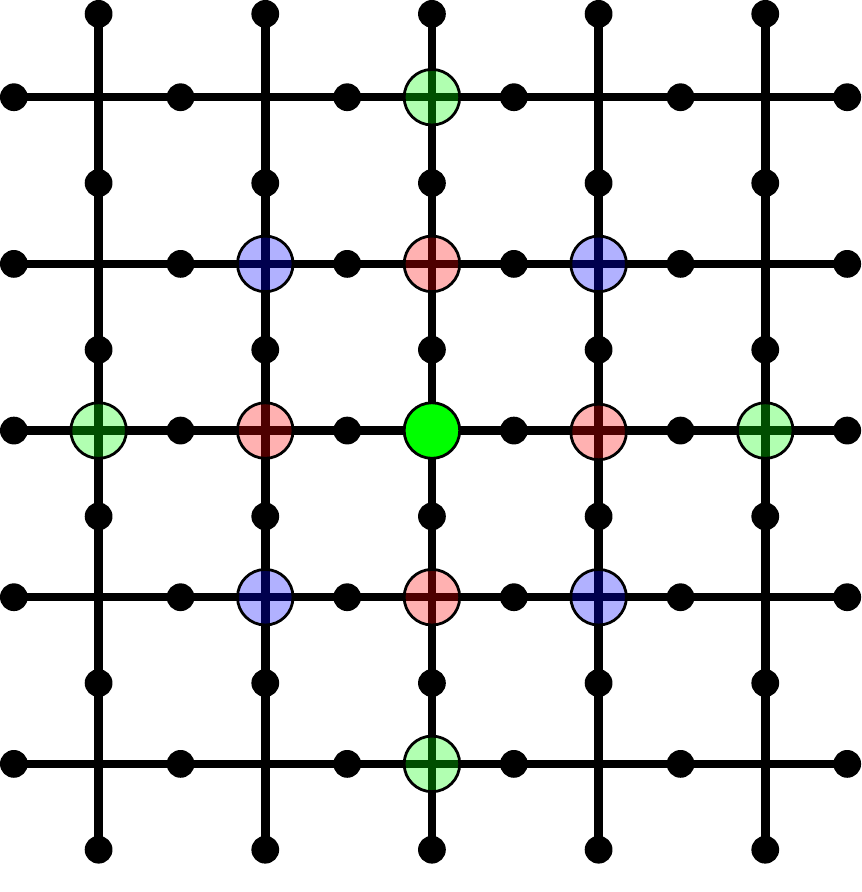}}
\end{center}
\caption{The eight starting geometries (translucent green, blue) for Monte Carlo simulations of random walks.  The solid green site denotes the origin at which the ``fixed'' quasi-particle sits.  Blue configurations were sampled twice as often as green configurations, owing to the different likelihoods of different starting geometries.  The simulation was terminated when the traveling quasiparticle reached one of the translucent red vertices---i.e. an annihilation geometry.}
\label{fig:ImpSampCartoon}
\end{figure}

In the low temperature limit, quasiparticle dynamics are dominated by trivial events where pairs that are created and immediately annihilate as $\gamma_- \gg \gamma_0$. Additionally, for nontrivial walks, once the anyon pair returns to occupy nearest neighbor sites on the lattice, the pair will annihilate with high probability. To model this low temperature regime, we consider an annihilation to occur as soon as a single random walker returns to a site adjacent to the origin \figref{fig:ImpSampCartoon}.  This can be understood as a ``zero temperature'' limit to the true quasiparticle statistics, as it ignores higher order processes that occur at finite temperature involving quasiparticle trajectories that meet in annihilation geometries, but then do not annihilate.  Explicitly, this approximation amounts to taking $\gamma_- \rightarrow \infty$.   Additionally, to improve efficiency of the Monte Carlo simulations, we start the walker at one of eight starting positions away from the origin; we account for the relative probabilities for reaching these starting positions via exact enumeration of the combinatorics of short topologically trivial walks (see \figref{fig:ImpSampCartoon}). The random walker undergoes a discrete time random walk on the square lattice on a torus until it returns to one of the four vertices adjacent to the origin. 

Using this approach, we compute the probability that two random walkers will annihilate after $n$ steps, $p^{\rm{t}} (n)$, and the average number of steps before annihilation, $\langle n \rangle$. For the true finite temperature toric code, the annihilation probability for nearest neighbor quasiparticles is less than one, as it is a function of $\gamma_0/\gamma_-$. The finite temperature probabilities $P^\Omega_{\rm{2D}}(L)$ may be computed from the zero temperature limit via a ``resummation" method described in Appendix \ref{sec:resum}.

\begin{figure}
\begin{center}
\scalebox{1}{\includegraphics[width=\columnwidth]{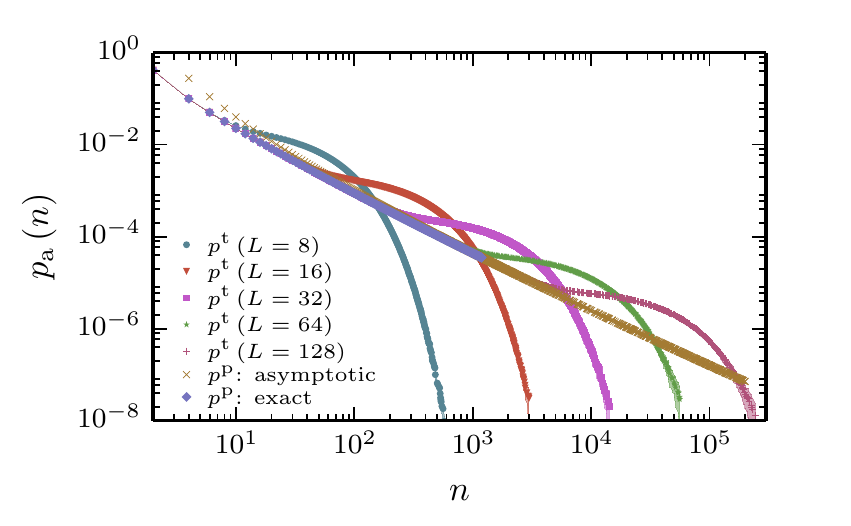}}
\end{center}
\caption{ Probability of annihilation $p^{\rm{t}} (n)$ after $n$ steps on the torus for the model described in section \ref{sec:MC2DRW}, as computed via Monte Carlo. We see that the value of $p^{\rm{t}} (n)$ agrees with the planar value $p^{\rm{p}} (n)$ (see \eqref{eq:ppln} and \cite{Montalenti2000}) up until a characteristic value of $n$ where it is possible for the walker to make topologically nontrivial walks on a finite size lattice.}
\label{fig:pa}
\end{figure}

\begin{figure}
\begin{center}
\scalebox{1}{\includegraphics[width=\columnwidth]{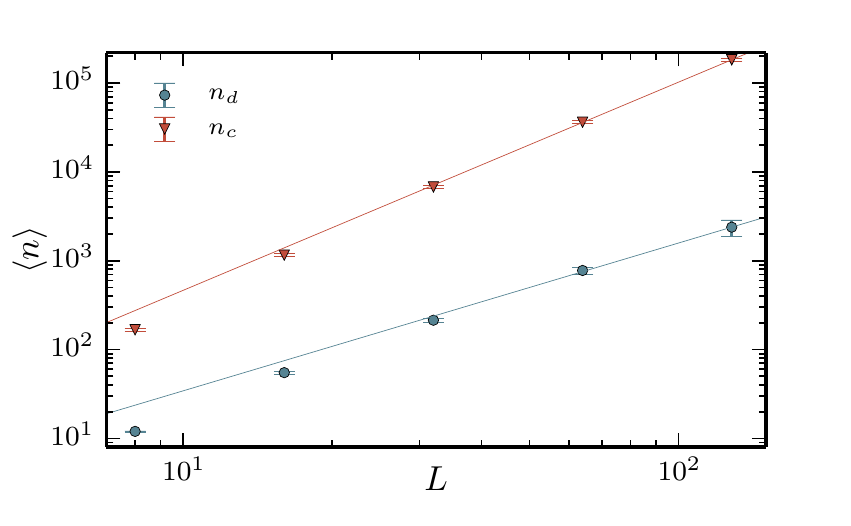}}
\end{center}
\caption{\label{fig:ndc} The finite-size scaling of the characteristic departure ($n_d$) and crossing ($n_c$) times from an analysis of the Monte Carlo data shown in \figref{fig:pa}. The lines represent the best fit power laws to the three largest system sizes.}
\end{figure}

\figref{fig:pa} shows the annihilation probability on a torus $p^{\rm{t}} (n)$ as a function of the number of steps $n$ for the zero temperature model with several system sizes $L$, as computed via Monte Carlo. We see that $p^{\rm{t}} (n)$ agrees with $p^{\rm{p}} (n)$ up to a certain value of $n$ for each lattice size. We can therefore define a characteristic ``departure time" $n_d(L)$,  as:
\begin{equation}
\left \vert \frac{p^{\rm{p}}\left(2n_d\right)-p^{\rm{t}}\left(2n_d \right)}{p^{\rm{p}}\left(2n_d\right)}\right \vert = \frac{1}{4}
\end{equation}
Random walks that annihilate at small $n$ are not sensitive to the topology of the finite-size lattice.  Thus, $n_{d}$  reflects the characteristic number of steps at which the random walk distribution is affected by the finite size torus topology. For $n>n_d$, we see that $p^{\rm{t}} (n) > p^{\rm{p}} (n)$ up to a characteristic number of steps.  We therefore define the ``crossing time" $n_c$ where $p^{\rm{t}} (n)$ crosses $p^{\rm{p}} (n)$ and then drops significantly. \figref{fig:ndc} shows the scaling of both dynamical quantities, $n_c$ and $n_d$ as a function of system size $L$.  Both are seen to be well described by power laws:
\begin{equation}
n_{c,d} \sim L^{\alpha_{c,d}} \label{eq:ncd}
\end{equation}
with $\alpha_c = 2.343\pm0.001$ and $\alpha_d = 1.66\pm0.04$.
 
\begin{figure}
\begin{center}
\scalebox{1}{\includegraphics{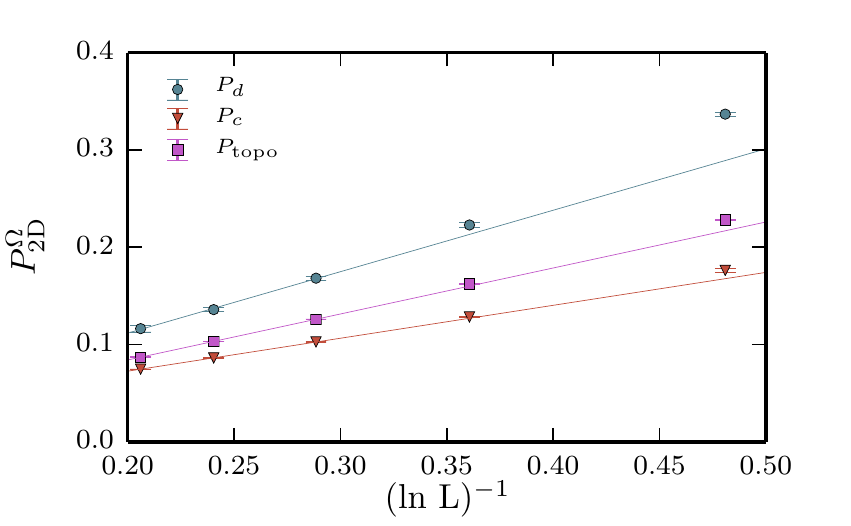}}
\end{center}
\caption{\label{fig:PvsL} The probability of topologically non-trivial annihilations on a torus, $P^{\Omega }_{{\rm 2D}}(L )$ as a function of system size, as computed by Monte Carlo simulations. Also shown are the bounds $P_c$ and $P_d$, which are computed from $p^{\rm{t}}(n)$ and $p^{\rm{p}}(n)$, as described in the text. The lines represent fits to $(\ln L)^{-1}$ of the three largest system sizes.}
\end{figure}

We also compute the initial and final topological sectors of each walk; this allows us to compute the probability of {\it topologically nontrivial annihilation}, $P^{\Omega }_{{\rm 2D}}(L )$, where the walk generates a topologically non-trivial path with odd winding. \figref{fig:PvsL} shows the finite size scaling of $P^{\Omega }_{{\rm 2D}}(L )$; for larger system sizes, we find
\begin{equation}
P^{\Omega }_{{\rm 2D}}(L )\approx \frac{c^{\Omega }_{{\rm 2D}}}{\ln(L)}
\end{equation}
where $c^{\Omega }_{{\rm 2D}} =  0.472\pm0.003$.

We may understand the scaling of $P^{\Omega }_{{\rm 2D}}(L )$ from an analysis of $p^{\rm{t}}(n)$. Since the deviations of $p^{\rm{t}}(n)$ from $p^{\rm{p}}(n)$ for $n > n_d$ are due to topologically nontrivial walks which contribute to $P^{\Omega }_{{\rm 2D}}(L )$, we can place approximate bounds on $P^{\Omega }_{{\rm 2D}}(L )$ from $p^{\rm{t}}(n)$. As an upper bound to $P^{\Omega }_{{\rm 2D}}(L )$, we assume that all walks for $n < n_d$ generate topologically nontrivial windings that contribute to $P^{\Omega }_{{\rm 2D}}(L )$; therefore we define the integrated probability:
\begin{align}
P_{d} \equiv \int^{\infty}_{n_d} dn~p^{\rm{p}} \left(n\right).
\end{align}
As both $p^{\rm{p}}$ and $p^{\rm{t}}$ integrate to unity, $P_d$ is approximately equal to the same integral over $p^{\rm{t}}$. $P_d$ includes both topologically trivial walks and topologically nontrivial walks that have even winding; consequently we expect $P_d$ to provide an upper bound to $P^{\Omega }_{{\rm 2D}}(L )$.

Alternately, we can make the approximation that topologically nontrivial walks are only the {\it excess} probability for $n_d \leq n \leq n_c$. By making the assumption that  $p^{\rm{t}}(n>n_c) \approx 0$ (see \figref{fig:pa}), we may approximate this excess by:
\begin{align}
P_{c} \equiv \int^{\infty}_{n_c} dn~p^{\rm{p}} \left(n\right),
\end{align}
again relying on the normalization of $p^{\rm{t}}$ and $p^{\rm{p}}$. While $P_c$ should over-count the events that contribute to $P^{\Omega }_{{\rm 2D}}(L )$ in the region $n_d \leq n \leq n_c$ (as only odd winding topological events contribute), the approximation $p^{\rm{t}}(n>n_c) \approx 0$ leads to an underestimation of $P^{\Omega }_{{\rm 2D}}(L )$, i.e. $P_c$ provides a lower bound on $P^{\Omega }_{{\rm 2D}}(L )$.

\figref{fig:PvsL} shows the finite size scaling of $P_c \left( L \right)$ and $P_d \left( L \right)$ and confirms that these provide a lower and upper bound to $P^{\Omega }_{{\rm 2D}}(L )$, respectively.  Thus as with $P^{\Omega }_{{\rm 2D}}(L )$, we find that $P_c \left( L \right)$ and $P_d \left( L \right)$ scale as $(\ln L)^{-1}$; the lines in \figref{fig:PvsL} represent a fit to $(\ln L)^{-1}$.

To see the origin of this $(\ln L)^{-1}$ scaling, we can use the asymptotic form of $p^{\rm{p}}(n)$ and the power law scaling of $n_{c,d}$ from \eqref{eq:ncd}, to approximate both $P_c$ and $P_d$:
\begin{align}
 P_{c,d} \left( L \right) &\approx \int^{\infty }_{n_{c,d}} dn~ \frac{1}{ n_{c,d}  \left( \ln n_{c,d} \right)^2 }=\frac{1}{\ln n_{c,d} } \notag \\
 & \approx  \frac{1}{\alpha_{c,d} \ln L }.
\end{align}
Consequently, we can understand the origin of the $(\ln L)^{-1}$ scaling which we predicted for $P^{\Omega }_{{\rm 2D}}(L )$ to fundamentally be due to the particular form of the polylog scaling of $p^{\rm{p}} (n)$. We note that this inverse logarithmic scaling of $P^{\Omega }_{{\rm 2D}}(L )$ implies a nontrivial finite-size scaling of ${\Gamma }_{\rm{TC}}$; indeed for the phenomenological form from Eq. \eqref{eq:GammaTC} we have ${\Gamma }_{\rm{TC}}\sim L^2/\ln L$.

\section{Real time Monte Carlo Simulation of the toric code dynamics}
\label{sec:MCmethod}

\subsection{Numerical method}

We now present Monte Carlo simulations of the real time dynamics of the toric code in contact with an Ohmic bath, as described in section \ref{sec:TCME}. We use a continuous real time Monte Carlo method~\cite{Chesi2010b} to numerically solve the master equation given in \eqref{eq:ME}. We focus on the relaxation dynamics of the system when prepared initially in a pure ground state. We define the operator:
\begin{equation}
\Pi_{++} \equiv \frac{1}{4}\left( W_1^z+1 \right) \left( W_2^z+1 \right);
\end{equation}
which is one for the $\vert \Psi_0^{++} \rangle$ ground state and vanishes for all other ground states. We then compute the expectation value $\langle \Pi_{++} \left( t \right) \rangle$ to study the decay from $\vert \Psi_0^{++} \rangle$. 

\begin{figure}
\begin{center}
\scalebox{1}{\includegraphics[width=\columnwidth]{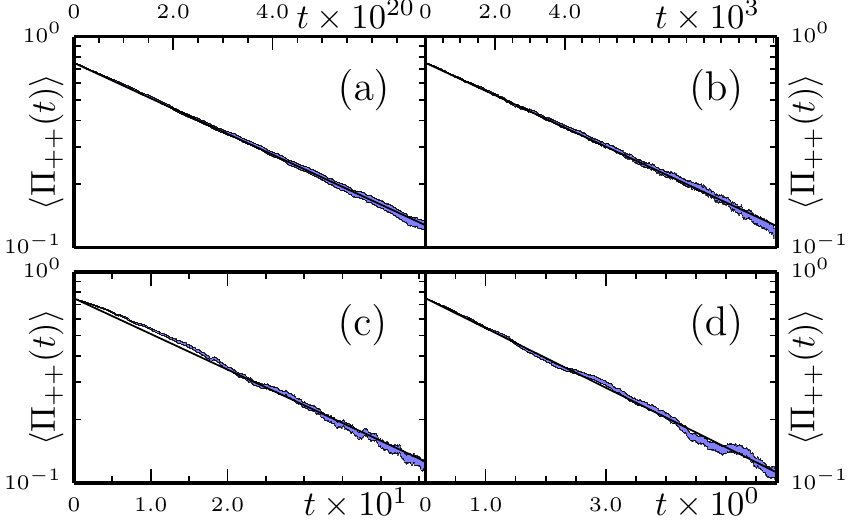}}
\end{center}
\caption{ Time evolution of the expectation value of the $\Pi_{++}$ operator (defined in section \ref{sec:MCmethod}) computed via Monte Carlo, where $1/4$ has been subtracted to reveal the exponential decay. These simulations were initialized to a pure ground state with $L=128$ with $T=\{ 0.02,0.08,0.14,0.2 \}$ respectively for subfigures (a)-(d). The black lines represent exponential fits to the Monte Carlo data. Note the stretched exponential behavior for early times in (c).  $\gamma_0$ has been set to 1.}\label{fig:decays}
\end{figure}

\begin{figure*}
\begin{center}
\subfloat{\scalebox{1}{\includegraphics[width=\columnwidth]{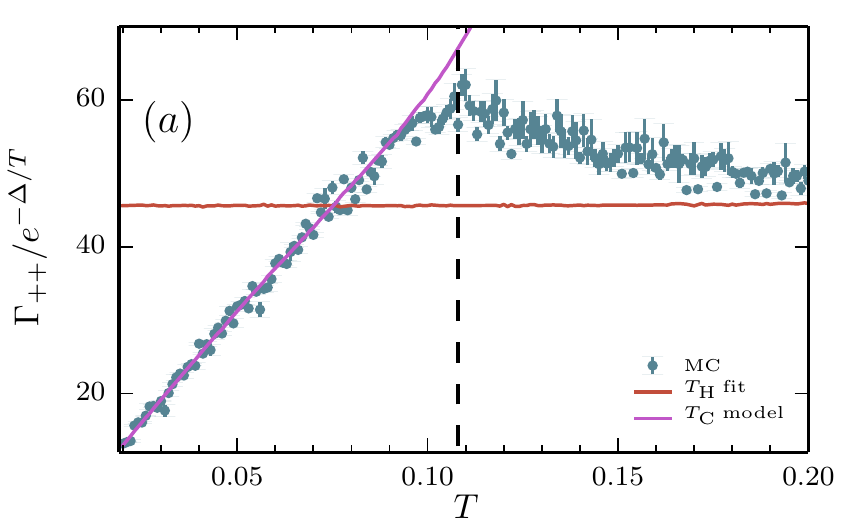}}}
\subfloat{\scalebox{1}{\includegraphics[width=\columnwidth]{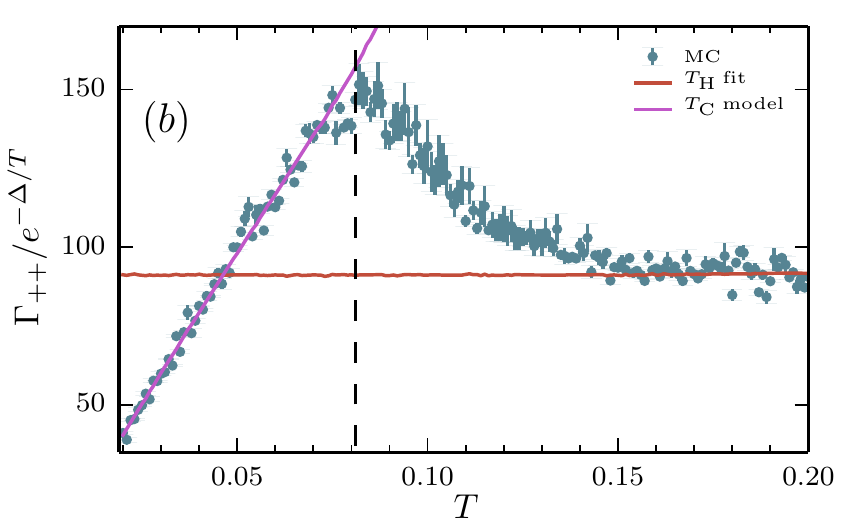}}}  
\\
\vspace{-1.2\baselineskip}
\subfloat{\scalebox{1}{\includegraphics[width=\columnwidth]{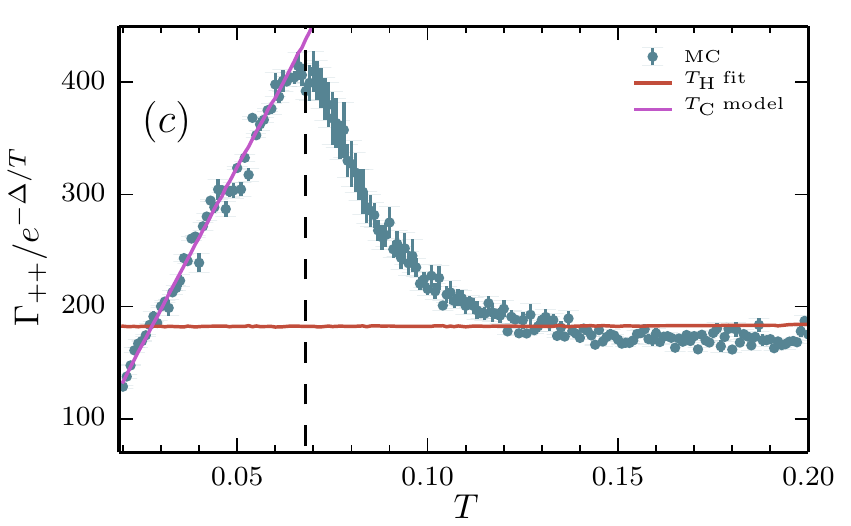}}}
\subfloat{\scalebox{1}{\includegraphics[width=\columnwidth]{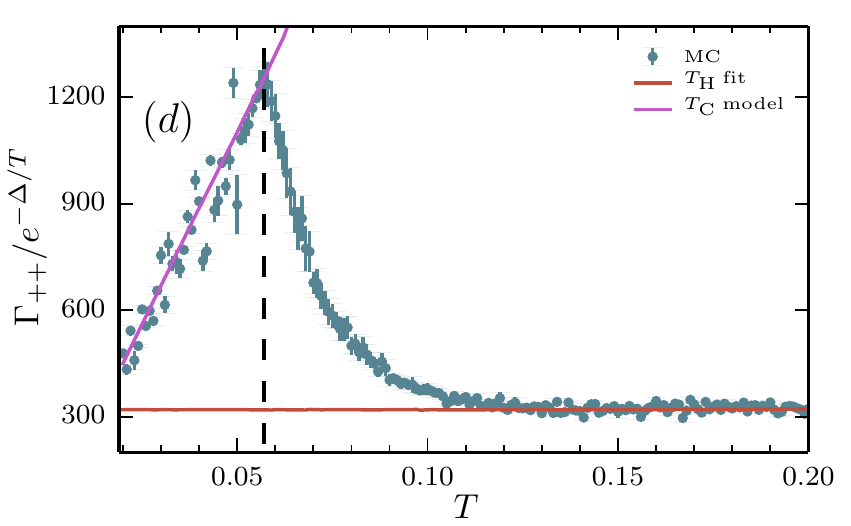}}}
\end{center}
\caption[justification=raggedright]{
\label{fig:Gamma00} Ground state relaxation rates for as a function of temperature for system sizes $L=\{16,32,64,128\}$ corresponding to (a)-(d) respectively. The solid lines are the low temperature phenomenological model, ($T_{\rm{C}}$), \eqref{eq:GammaTC}, and the high temperature fit, $T_{\rm{H}}$, \eqref{eq:GammaTH}. The vertical dashed line represents the dynamical crossover temperature $T^*_{\rm{dyn}}$. Note that $\Gamma_{++}(T)$ is a monotonic, increasing function of $T$; the rescaling by $\exp({-\Delta/T})$ generates the nonmonotonicity.  $\Delta$ has been set to 1.
}
\label{fig:GammavsT}
\end{figure*}

The exponential nature of the decay of $\Pi_{++}(t)$ is displayed in \figref{fig:decays}, where the lines represent exponential fits to the Monte Carlo data.  We find such decays are well described by exponential decays at all but intermediate temperatures (see \figref{fig:decays} c.), where short time deviations lead to a stretched exponential decay.  We fit $\Pi_{++}(t)$ to an exponential:
\begin{equation}
\Pi_{++}\left( t \right ) = \frac{1}{4}\left(1+3e^{-\Gamma_{++}t} \right)
\end{equation}
 to extract the relaxation rate $\Gamma_{++}$.  The additional systematic uncertainty of the rate $\Gamma_{++}$ due to the stretched exponential behavior in intermediate regimes does not appreciably affect the analysis.
 \figref{fig:GammavsT} shows $\Gamma_{++}/e^{-\Delta/T}$ computed for four system sizes over a range of temperatures. We see three temperature regimes for each system size: a low temperature regime where $\Gamma_{++} \sim T e^{-\Delta/T}$, a high temperature regime where $\Gamma_{++} \sim e^{-\Delta/T}$ and an intermediate temperature regime smoothly connecting these two forms of the temperature scaling.

\subsection{Low Temperature regime}

\begin{figure}
\begin{center}
\scalebox{1}{\includegraphics[width=\columnwidth]{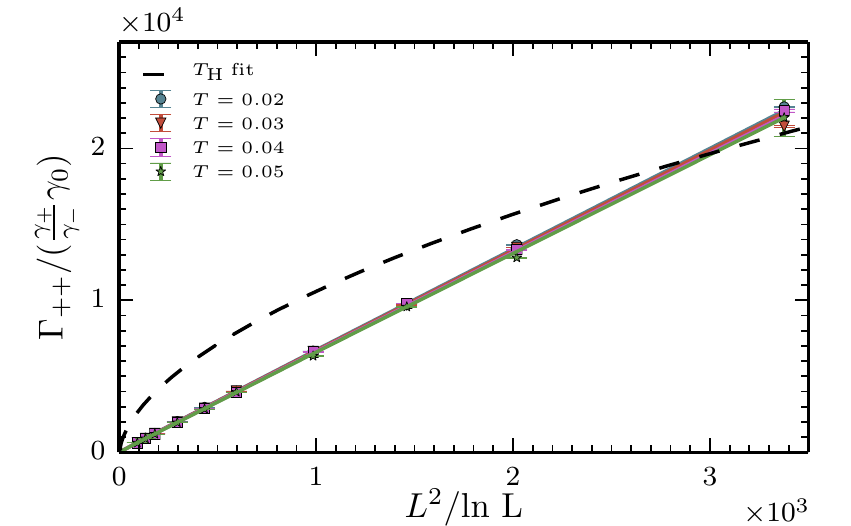}}
\end{center}
\caption{Low temperature regime: Finite-size scaling of $\Gamma_{++}$ in the low temperature regime where we have collapsed the temperature dependence of the data according to \eqref{eq:GammaTC}. The solid lines are the low temperature model ($T$) predictions as described in \eqref{eq:Markov}; these lines nearly completely overlap due to the weak residual temperature dependence in \eqref{eq:GammaTC}.  The dotted line indicates the best fit to purely linear scaling in $L$ which is expected in the high temperature regime ($T_{\rm{H}}$).}
\label{fig:LowTCollapsed}
\end{figure}

 \figref{fig:GammavsT} shows the low temperature model predictions of  \eqref{eq:GammaTC} as well as the Monte Carlo data. The regime of linear behavior of $\Gamma_{++}/e^{-\Delta/T}$ and corresponding agreement with the effective model at low temperatures allows us to identify this regime as the low temperature regime where the finite-size scaling of the relaxation time is determined by the scaling of topologically non-trivial random walks. \figref{fig:LowTCollapsed} shows the finite-size scaling of $\Gamma_{++}$ in this low temperature regime, where we have performed a data collapse to remove the leading temperature dependence of \eqref{eq:Markov}. We find good agreement with the \emph{parameter free} low temperature model \eqref{eq:Markov} (with temperature dependent $P^\Omega_{\rm{2D}}(L)$ obtained by the resummation procedure in appendix A) which has an approximate $L^2/\ln L$ scaling. This non-trivial finite-size scaling is a key feature of this low temperature regime. 

On increasing temperature, we can understand the transition out of this low temperature regime as follows. The separation of timescales breaks down as the ratio $L^2\cdot\gamma_{+}/\gamma_{0}$ grows larger. As the temperature increases, the decay rate is significantly affected by multi-pair processes which are not accounted for in \eqref{eq:Markov}.
Interactions between multiple pairs of quasiparticles modify the annihilation probability distributions used in the low temperature model. Additionally, at higher temperatures, as the lifetime of quasiparticle pairs increases, the decay of $\Pi_{++}$ is sensitive to trivial error strings (i.e., single applications of $E^{e\dagger}$) acting across the edges shared with the winding operators. At higher temperatures, these trivial error strings dominate the decay rate.

\subsection{High temperature regime}

\begin{figure}
\begin{center}
\scalebox{1}{\includegraphics[width=\columnwidth]{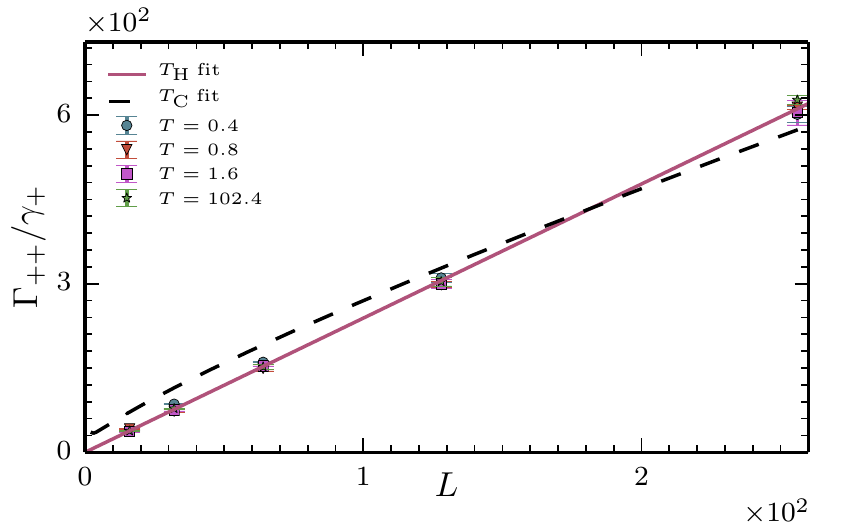}}
\end{center}
\caption{High temperature regime: finite-size scaling of $\Gamma_{++}$ in the high temperature regime. The solid line represents a fit of all high temperature data ($T_{\rm{H}}$) to a linear scaling in $L$.  The dotted line indicates a best fit to the poly-log scaling $L^2/\rm{log} L$ which is expected in the low temperature regime ($T_{\rm{C}}$).}
\label{fig:HighTCollapsed}
\end{figure}

At higher temperatures in \figref{fig:GammavsT}, we see that $\Gamma_{++} \sim e^{-\Delta/T}$; this is the high temperature regime described by \eqref{eq:GammaTH} where we expect a linear scaling in system size. In \figref{fig:HighTCollapsed} we show a fit to the linear finite-size scaling for several different temperatures, where we have scaled the $\Gamma_{++}$ by the rate of formation of quasiparticle pairs, $\gamma_+$. This one parameter linear fit to the scaled data gives a single functional form for all system sizes and temperatures 
\begin{align}
{\Gamma }_{++} = c_{T_H} \gamma_+ L,
\end{align}
where we find the constant $c_{T_H} = 2.5 \pm 0.1$. If only the lowest order process trivial anyon pairs contributed to the decay across both winding operators, we would have $c_{T_H} = 2$.  Obtaining a fit to the finite-size scaling with $c_{T_H} >2$ suggests that higher order processes are also providing significant contributions. The solid red line in \figref{fig:GammavsT} shows that this one parameter high temperature fit is in good agreement with the Monte Carlo data in the high temperature regime. We note that the linear finite-size scaling of $\Gamma_{++}$ is distinct from the $L^2 / \ln{L}$ scaling in the low temperature regime (see \figref{fig:LowTCollapsed}).
 
\subsection{Dynamical Crossover Temperature}

\begin{figure}
\begin{center}
\scalebox{1}{\includegraphics[width=\columnwidth]{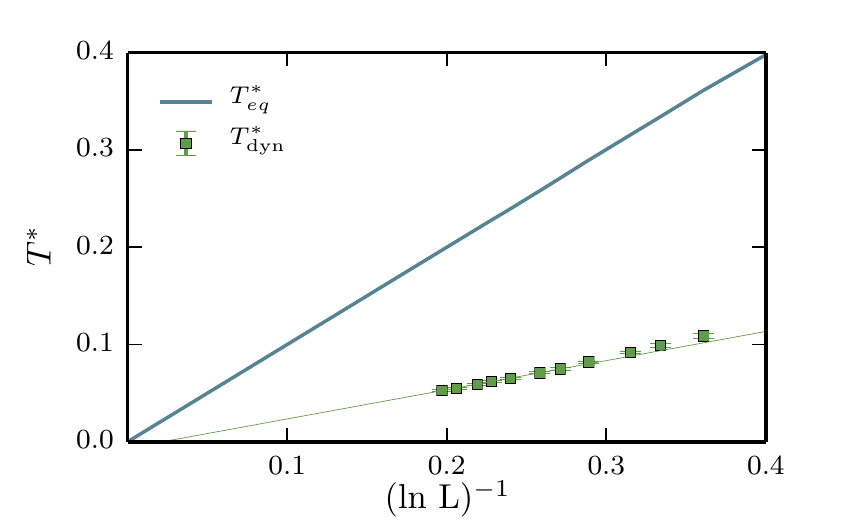}}
\end{center}
\caption{The dynamical crossover temperature $T^*_{\rm{dyn}}$ as a function of system size $L$. The line represents the fit to $\ln(L)^{-1}$ scaling for the largest system sizes. Also shown is the equilibrium crossover temperature $T^*_{\rm{eq}}$ as defined in Ref. \onlinecite{Castelnovo2007a}.}
\label{fig:T*dyn}
\end{figure}

Analysis of the results shown in \figref{fig:GammavsT} strongly suggests that we can identify two distinct regimes where the relaxation rate is dominated by distinct physical processes. We can therefore define a {\it dynamical crossover temperature} $T^*_{\rm{dyn}}$ which signifies the crossover between these regimes. We define $T^*_{\rm{dyn}}$ as the local maxima on \figref{fig:GammavsT} where the linear temperature scaling breaks down; this does not correspond to a maximum of $\Gamma_{++}$ itself, which is monotonically increasing as a function of temperature, since we have removed the temperature dependence of the Boltzmann factor by rescaling. Clearly $T^*_{\rm{dyn}}$ is a function of system size, since the low-temperature regime shrinks as $L$ increases; \figref{fig:T*dyn} displays the finite-size scaling of $T^*_{\rm{dyn}}$ as well as the equilibrium crossover temperature $T^*_{\rm{eq}}$ computed in [\onlinecite{Castelnovo2007a}] from the topological entanglement entropy. We find an inverse logarithmic scaling of $T^*_{\rm{dyn}}$ with system-size, in agreement with the scaling of $T^*_{\rm{eq}}$.
 
\section{Discussion}

We have demonstrated the non-trivial finite size scaling of the relaxation time of the toric code in contact with a thermal reservoir, using numerical simulations of real time dynamics of quasiparticles. We have identified a low temperature regime in which the relaxation dynamics are dominated by topologically non-trivial random walks of quasiparticle pairs; consequently the finite-size and temperature scaling of this regime are distinct from the high temperature regime above the crossover temperature. We find that both the finite-size and finite temperature scaling are \emph{stronger} in this low temperature regime than at higher temperatures where the behavior coincides with the expected scaling in the thermodynamic limit\cite{Viyuela2012}.  

In the low temperature regime, we find the relaxation rate to scale as $L^2/\ln L$, in contrast to the scaling as $L$ above $T^*_{dyn}$. Consequently, the lifetime of topological qubits will increase faster in this regime as the system size is decreased, than above $T^*_{dyn}$. We also find that the relaxation rate is suppressed by an additional factor of $T$ in the low temperature regime; the memory lifetime will increase with inverse temperature $\beta=1/T$  as $\beta e^{\Delta \beta}$, faster than the $e^{\Delta \beta}$ scaling of the lifetime above $T^*$. We note that the particular form of the additional crossover suppression is dependent on the nature of the bath, since it arises from the temperature scaling of the diffusion rate for the ohmic bath studied here, $\gamma_0 = \xi k T$.  In contrast, for a super-Ohmic bath $\gamma_0 = 0$; however, the effective diffusion rate will scale as $e^{-2\Delta/T}$, due to indirect hopping of quasiparticles from 2nd order pair creation events~\cite{Chesi2010a}. Consequently the low temperature suppression will be even stronger for a super-Ohmic bath.

This work may help guide the design of ground state relaxation of optimal topological qubits. While it is now evident that true topological protection is not achievable for the 2D toric code in the thermodynamic limit, as a practical matter in a finite size realization, one may wish to balance robustness to unitary perturbations, which is maximized by using the largest possible system, against thermal robustness, which decreases with system size. The stronger finite-size and temperature scaling of the relaxation time (corresponding to the quantum memory lifetime) in the low temperature regime  suggests that the optimal balance will be achieved below $T^*_ {\rm{dyn}}$.  The corresponding optimal size will of course depend on the prefactors of the scaling relations and will therefore be dependent on 
both the microscopic form of the coupling to the bath 
and the unitary perturbations.  

Thus, although the topological order required for topological protection of quantum information processing is destroyed at all temperatures in the thermodynamic limit, we have identified a dynamical low temperature regime for finite size systems which may prove practically useful for quantum information processing.

\section{Acknowledgments}
This material is based upon work supported by DARPA under Grant No. 3854-UCB-AFOSR-0041 and by NSF under Grant PIF-0803429.  CDF was supported by the NSF Graduate Research Fellowship under Grant DGE-1106400.  CDF also thanks Stefano Chesi for helpful correspondence regarding bath models.

\appendix

\section{Resummation method}

\label{sec:resum}

 The probabilities of return were calculated by numerically tabulating the fraction of random walks that arrived at the annihilation geometries depicted in \figref{fig:ImpSampCartoon}. At finite temperature, quasiparticles have a nonzero probability of not annihilating after reaching these positions, and continuing a random walk.  Only in the `zero temperature' limit do these quantities represent the true annihilation statistics for the quasiparticles.  To distinguish between these, we define $\overline{P^{\Omega}_{\delta 1}}$ and $\overline{P^{\Omega}_{\delta 2}}$ as the ``zero temperature'' probabilities of return.
 
 To calculate this temperature dependent annihilation probability, we define:

 \begin{equation}
\overline{P_{ij}^{\Omega}} = 
\begin{cases} \overline{P^{\Omega}_{\delta 1}} &\mbox{if } i,j  \mbox{ differ by one axis } \\ 
\overline{P^{\Omega}_{\delta 2}} &\mbox{if } i,j  \mbox{ differ by both axes } \\
1 - 2 \overline{P^{\Omega}_{\delta 1}} - \overline{P^{\Omega}_{\delta 2}} &\mbox{if } i=j  
\end{cases} 
\end{equation}
 
 $\overline{P_{ij}^{\Omega}}$ represents the transition matrix for a \emph{discrete} Markov chain.  This matrix encodes the zero temperature transit probabilities for a quasiparticle pair to meet in an annihilation geometry.  To account for the possibility of both annihilation and continued traversal, we define:
\begin{align}
 \Sigma =\left( \begin{array}{cc}
(1-\tau )\overline{P_{ij}^{\Omega}} & \mathbf{0} \\ 
\tau\overline{P_{ij}^{\Omega}} & {\mathbb I} \end{array}
\right), \\
\tau = \left(\frac{\gamma_-}{6 \gamma_0 +\left(2 L^2-7\right)\gamma_+ + \gamma_-}\right),
\end{align}
where $\tau $ is the probability that an adjacent pair of quasiparticles annihilates, ${\mathbb I}$ is the $4\times4$ identity matrix, and where $\mathbf{0}$ represents a $4\times4$ zero matrix.  The initial state vector for this Markov chain represents a single pair of quasiparticles initialized to one of the starting configurations in a given sector.  By convention, these are the $++, +-, -+, --$ sectors for the first four entries of the state vector.  The latter four entries encode the probabilities of a pair of walkers annihilating in a given sector after some number of steps.  The long time steady state solution of this larger Markov chain then determines the temperature dependent probabilities that a given quasiparticle pair causes a transition.

For example, consider a pair initialized to the $++$ sector, with an initial state vector $\left(1,0,0,0,0,0,0,0\right)^T$. The state in the long time limit is
\begin{align}
\lim_{k \to \infty} \Sigma^k \cdot &\left( 1,0,0,0,0,0,0,0 \right)^T = \notag \\ 
&\left( 0,0,0,0,1-2P^{\Omega}_{\delta 1}-P^{\Omega}_{\delta 2},P^{\Omega}_{\delta 1},P^{\Omega}_{\delta 1},P^{\Omega}_{\delta 2} \right)^T
\end{align}
In the higher temperature limit all transition probabilities tend towards $1/4$.  The zero temperature limit corresponds to the bare probabilities $\overline{P^{\Omega}_{\delta 1}}$ and $\overline{P^{\Omega}_{\delta 2}}$. The temperature dependent $P^{\Omega}_{\delta 1,2}$ are used in the manuscript in sections III and V.  The ``zero temperature'' limits are used exclusively in section IV.

\section{Finite size scaling in the 1D Ising model}
\label{sec:P1D}

\begin{figure}
\includegraphics*[width=0.75\columnwidth]{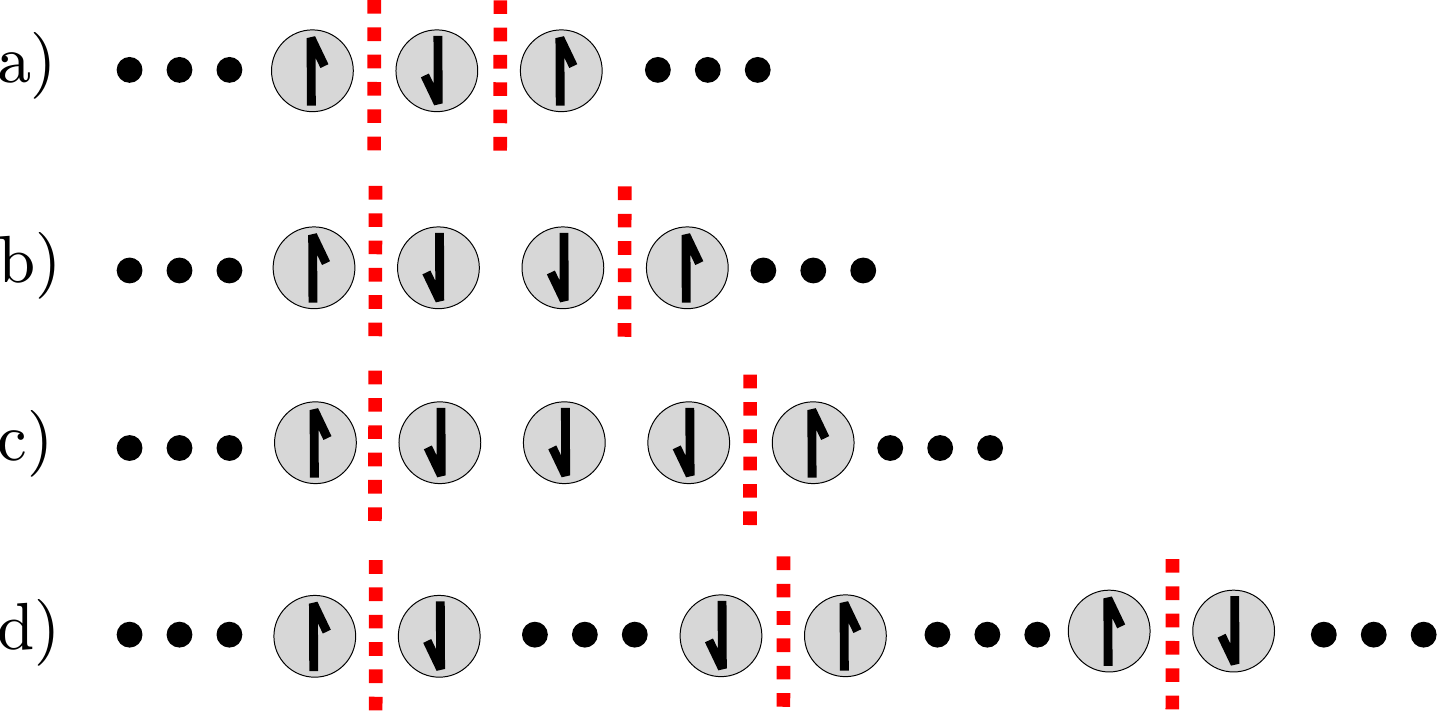}
\caption{Various configurations of domain walls in 1 dimension.  a) An ``annihilation geometry'', where the domain walls (in red) are separated by a single spin.  b) A ``free'' pair of domain walls.  This configuration will annihilate trivially with probability 1/2.  c) Domain walls are further separated. d) Domain walls are separated by half the system size.  The rightmost domain wall is the same as the leftmost domain wall due to periodic boundary conditions.}
\label{fig:po1d}
\end{figure}

Here we demonstrate the scaling of $P^\Omega_{1D}$ discussed in \ref{sec:Ising}. Consider a pair of domain walls on a 1D periodic chain of $L$ classical Ising spins separated by two spins (i.e., configuration (b) in \figref{fig:po1d}). In such a configuration, it is equally likely for the domain walls to move one unit to the left or right.  If either of the domain walls is separated from the other by only a single spin, they are in an ``annihilation geometry'' (i.e., configuration (a) in \figref{fig:po1d}), and in the $T\rightarrow0$ limit will annihilate with unit probability.

Without loss of generality, fix one domain wall as an ``origin''.  A pair initially in configuration (b) from \figref{fig:po1d} will either annihilate with probability 1/2, or become separated by at least 2+1 spins with probability 1/2.  In random walks for which the domain walls become separated by 2+1 spins (i.e., configuration (c) in \figref{fig:po1d}), the domain walls will either annihilate trivially with conditional probability 1/2 or become separated by at least 4+1 spins with conditional probability 1/2.

In this way, the set of random walks available to a domain wall pair separated by $d+1$ spins can always be partitioned into those that return to the annihilation geometry, and those that separate the domain wall pair by an additional $d$ spins, because the inverse process that results in an annihilation event is a random walk that separates the domain walls by an additional $d$ spins.  Once $d+1$ is exactly half the system size, the probability of the domain walls separating by an additional $d$ spins is equivalent to annihilating nontrivially, as the free domain wall ``wraps around'' and annihilates from the opposite side of the fixed domain wall.

For simplicity, if we suppose the system size is of the form $L = 2^n + 2$ for some positive integer $n$, then domain walls which are separated by $L/2$ spins (equivalently, $2^{n-1}+1$ spins) have a conditional probability of 1/2 of annihilating either trivially or nontrivially.  The total probability of the domain walls reaching this configuration is just the product of the conditional probabilities of the domain walls reaching $2+1$, $4+1$, ..., $2^{n-1}+1$ spins of separation.  Thus: $P^\Omega_{1D} = 1/2^n$, or by rearrangement: $P^\Omega_{1D} = 1/(L-2)$.

\bibliography{manuscriptnewer}

\end{document}